\documentclass{Interspeech}
\usepackage{todonotes}
\usepackage{url}
\usepackage{CJKutf8}
\interspeechcameraready

\title{Mitigating Language Mismatch in SSL-Based Speaker Anonymization}

\author[affiliation={1}]{Zhe}{Zhang}
\author[affiliation={2}]{Wen-Chin}{Huang}
\author[affiliation={1}]{Xin}{Wang}
\author[affiliation={3}]{Xiaoxiao}{Miao}
\author[affiliation={1}]{Junichi}{Yamagishi}

\affiliation{}{National Institute of Informatics}{Japan}
\affiliation{}{Nagoya University}{Japan}
\affiliation{}{Duke Kunshan University}{China}
\email{\{zhe,wangxin,jyamagis\}@nii.ac.jp, wen.chinhuang@g.sp.m.is.nagoya-u.ac.jp, xiaoxiao.miao@dukekunshan.edu.cn}

\usepackage{comment}
\usepackage{multirow}

\begin{document}

\maketitle

\begin{abstract}
    
Speaker anonymization aims to protect speaker identity while preserving content information and the intelligibility of speech. However, most speaker anonymization systems (SASs) are developed and evaluated using only English, resulting in degraded utility for other languages. This paper investigates language mismatch in SASs for Japanese and Mandarin speech. First, we fine-tune a self-supervised learning (SSL)-based content encoder with Japanese speech to verify effective language adaptation. Then, we propose fine-tuning a multilingual SSL model with Japanese speech and evaluating the SAS in Japanese and Mandarin. Downstream experiments show that fine-tuning an English-only SSL model with the target language enhances intelligibility while maintaining privacy and that multilingual SSL further extends SASs' utility across different languages. These findings highlight the importance of language adaptation and multilingual pre-training of SSLs for robust multilingual speaker anonymization.
\end{abstract}

\section{Introduction}

\begin{figure}[t]
  \centering
  \includegraphics[width=\linewidth]{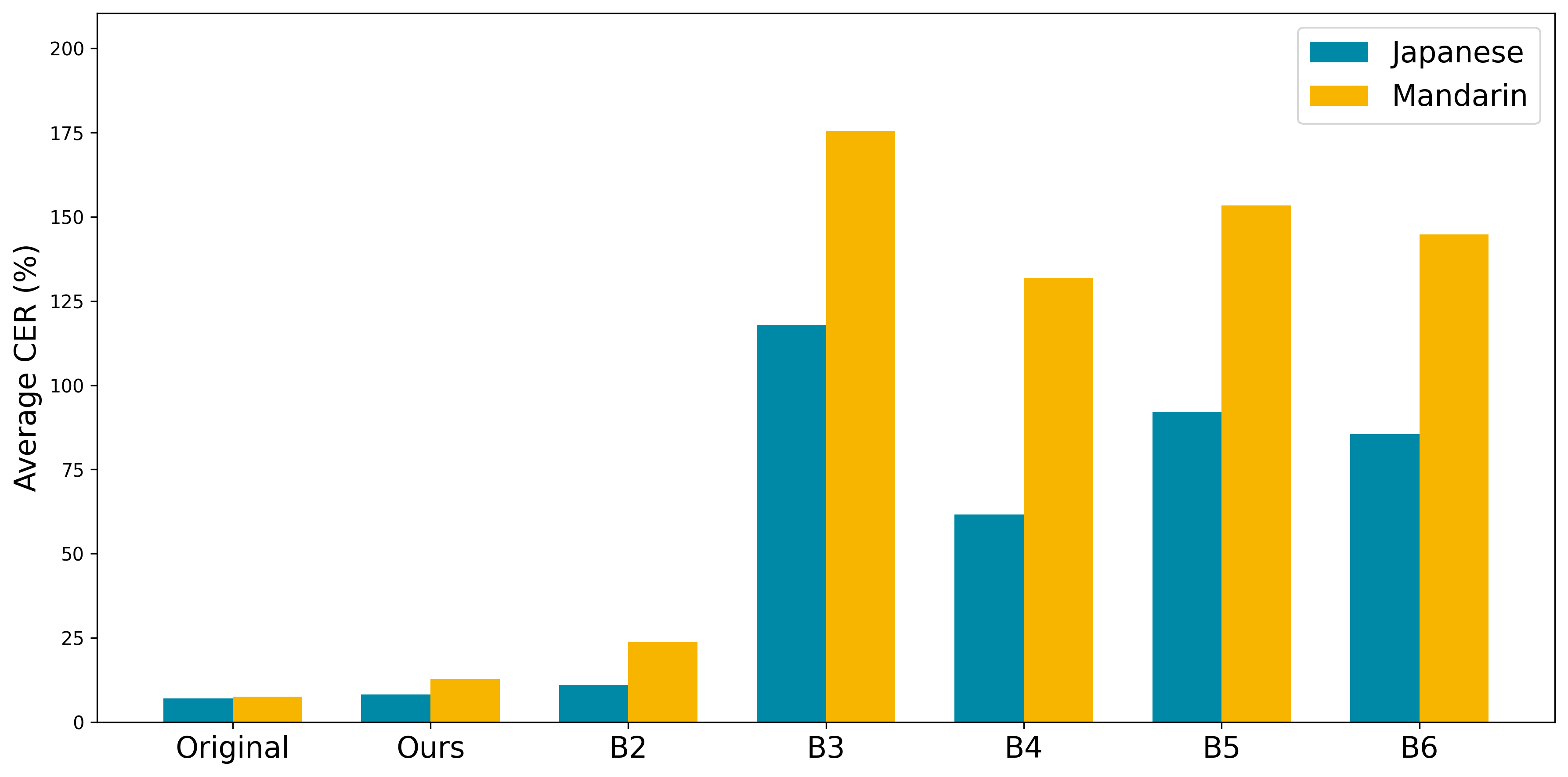}
  \caption{Utility evaluation via ASR task on anonymized Japanese and Mandarin speech using VPC baselines (B2 - B6).}
  \label{fig:baseline_cer}
\end{figure}

Speaker anonymization has gained increasing attention as privacy concerns grow in speech applications. With the rapid advancement of AI-driven speech technologies, it is crucial to ensure that sensitive speaker characteristics remain concealed while preserving speech's linguistic content and naturalness. Recent advances, as demonstrated in the VoicePrivacy Challenges (VPC) \cite{tomashenkoIntroducingVoicePrivacyInitiative2020,tomashenkoVoicePrivacy2022Challenge2022,tomashenkoVoicePrivacy2024Challenge2024}, have led to promising approaches based on digital signal processing (DSP) and data-driven methods. VPC provides a unified evaluation framework to advance the development of voice privacy preservation techniques. However, most existing data-driven models are trained and tested primarily in English. This narrow linguistic scope raises questions about their generalization and robustness: \textit{What happens if non-English speech, such as Japanese and Mandarin, is processed by these SASs?} Our preliminary experiments with VPC baselines (B2 - B6) demonstrated a significant degradation of intelligibility for Japanese and Mandarin inputs, as measured by increased character error rates (CERs) in automatic speech recognition (ASR) tasks. This suggests that current SASs may not be robust to multilingual inputs, as shown in \autoref{fig:baseline_cer}.

The anonymized speech from the baseline models is unintelligible and unusable for downstream tasks. Some CER values can be even larger than 100\% because the ASR model cannot recognize the target language and thus transcribe alphabets, resulting in a large increase in insertion errors, demonstrating poor intelligibility. Although B2 shows a lower CER than others, it suffers a poor privacy preservation performance with a low equal error rate (EER) of 4.52\% in the automatic speaker verify (ASV) experiments \cite{tomashenkoVoicePrivacy2024Challenge2024}. Improving the utility of SASs while maintaining privacy is an important topic, especially for extending the usage of English-only models to different languages. Among the various SAS methods, the self-supervised learning (SSL)-based methods are expected to achieve speaker anonymization independent of language with higher accuracy. However, as seen from the CER values (e.g., B5), there is still a problem with low performance in multilingual settings.

We assume that this problem is caused by language mismatch; the SSL model used in SAS is trained only with English data, so it is less effective for other languages. Although papers have recently been published that address the issue of language mismatch in multilingual speaker anonymization systems \cite{meyerProbingFeasibilityMultilingual2024,yaoMUSAMultilingualSpeaker2024,miaoLanguageIndependentSpeakerAnonymization2022,miaoAnalyzingLanguageIndependentSpeaker2022}, ASR experiments still suffer from high CERs, which hinders practicality in downstream tasks. In this paper, we investigate multilingual SSL-based content encoders to solve the problem of language mismatch in speaker anonymization. Specifically, we consider the following two scenarios for using multilingual SSL-based soft content encoders:

\begin{enumerate}
    \item Fine-tuning a multilingual HuBERT model, mHuBERT \cite{zanonboitoMHuBERT147CompactMultilingual2024}, with a specific language, i.e., Japanese, to investigate whether the language-adapted SSL model is helpful for adapting the SAS to a particular language.

    \item Use the above multilingual HuBERT model fine-tuned to the specific language and examine whether it also contains meaningful representations for other languages, such as Mandarin.
\end{enumerate}

This is an essential step towards properly understanding the root of the language mismatch problem in multilingual speaker anonymization systems and proposing a solution. Note that other types of speech SSL models, such as Wav2Vec2 \cite{baevskiWav2vec20Framework2020} and XLS-R \cite{babuXLSRSelfsupervisedCrosslingual2022}, have also been studied in previous research \cite{tomashenkoVoicePrivacy2024Challenge2024,yaoMUSAMultilingualSpeaker2024}, but to eliminate the impact of different SSL architectures, we have limited our experiments to HuBERT-based models.

The main contributions of this paper are as follows:
\begin{itemize}
    \item We investigated the problems that occur when Japanese speech is input into SASs based on the English-only HuBERT model and analyzed how the problems change by fine-tuning it with Japanese speech data.
    \item We integrate a state-of-the-art mHuBERT model into the soft content encoder, achieving better performance in both Japanese and Mandarin, demonstrating that a multilingual SSL model will also benefit a broader linguistic scope.
    \item Phonetic analysis of Japanese speech demonstrates that the multilingual SSL model effectively improves the intelligibility of syllables difficult for ASR models to recognize.
\end{itemize}

Our findings underscore the importance of adapting SSLs to the target language and incorporating multilingual SSL models, paving the way for multilingual speaker anonymization. Audio samples are available on the demo page \footnote{\url{https://nii-yamagishilab.github.io/multilingual-SSL-SAS-samples/}}. We released the pre-trained weights and codes \footnote{\url{https://github.com/nii-yamagishilab/multilingual-SSL-SAS}}.

\section{Related Work}

\subsection{English-Based SASs}

Speaker anonymization is a voice privacy solution to conceal speaker identity without degrading intelligibility and naturalness \cite{tomashenkoIntroducingVoicePrivacyInitiative2020}. Aiming at standardizing and advancing the development of voice privacy preservation techniques, the VoicePrivacy Challenge (VPC) was initiated \cite{tomashenkoIntroducingVoicePrivacyInitiative2020} and held in 2022 \cite{tomashenkoVoicePrivacy2022Challenge2022} and 2024 \cite{tomashenkoVoicePrivacy2024Challenge2024}, instrumental in benchmarking anonymization methods. Traditional approaches to protecting speaker privacy are based on DSP techniques. McAdams coefficients are utilized to shift the formant positions of the original speaker \cite{patinoSpeakerAnonymisationUsing2021}. Although DSP algorithms are language-independent, they often suffer from low synthesis quality. State-of-the-art anonymization approaches adopt methods from voice conversion and synthesis, aiming to disentangle a speech utterance into speaker, content, and prosodic representations. By anonymizing speaker representation while keeping the original content and prosody, the anonymized speech can suppress the original speaker identity while maintaining intelligibility. Anonymization based on x-vector \cite{fangSpeakerAnonymizationUsing2019} is a typical approach used in the VPC baselines \cite{tomashenkoIntroducingVoicePrivacyInitiative2020,tomashenkoVoicePrivacy2020Challenge2022,tomashenkoVoicePrivacy2024Challenge2024}. It extracts speaker representations via a TDNN-based ASV system \cite{snyderXVectorsRobustDNN2018} and uses a selection-based speaker anonymizer \cite{srivastavaDesignChoicesXVector2020} to replace the original x-vector with the averaged vector from an external speaker pool. Based on this framework, some later works focused on improving speech disentanglement \cite{mawalimSpeakerAnonymizationModifying2022, pierreAreDisentangledRepresentations2022, shahinshamsabadiDifferentiallyPrivateSpeaker2023} to remove the speaker information in linguistic features. Other subsequent works mainly focused on improving the anonymization methods of the speaker vectors \cite{perero-codoseroXvectorAnonymizationUsing2022, turnerGeneratingIdentitiesMixture2022, meyerAnonymizingSpeechGenerative2022, chenSystemDescriptionVoice2022}. Such approaches use an automatic speech recognition neural acoustic model (ASR AM) to extract linguistic features. However, an ASR AM requires large amounts of transcribed training data, which are mainly limited to English, and thus can be ineffective for unseen languages. 

\subsection{Multilingual SASs}
How to enable multilingual speaker anonymization becomes a challenging topic for generalization of SASs. Multilingual ASR models are adopted as content encoders in \cite{meyerProbingFeasibilityMultilingual2024}. Meanwhile, SSL has emerged as a powerful paradigm for representation learning, showing remarkable performance for speech synthesis \cite{huangAnytoOneSequencetoSequenceVoice2021,polyakSpeechResynthesisDiscrete2021}. XLS-R \cite{babuXLSRSelfsupervisedCrosslingual2022} is used as a teacher model in distillation in \cite{yaoMUSAMultilingualSpeaker2024}. A language-independent speaker anonymization approach \cite{miaoLanguageIndependentSpeakerAnonymization2022} adopts an SSL-based soft content encoder, which demonstrates better performance on unseen language. A following analysis \cite{miaoAnalyzingLanguageIndependentSpeaker2022} discussed the language mismatch of SSL-based SASs in detail. A later work proposed an anonymization approach based on an orthogonal Householder neural network (OHNN), further alleviating the limitation of selection-based anonymizers \cite{miaoSpeakerAnonymizationUsing2023}. In these works, HuBERT-based content encoders \cite{vanniekerkComparisonDiscreteSoft2022} are adopted. However, the HuBERT model \cite{hsuHuBERTSelfSupervisedSpeech2021} is trained exclusively using English data, which still limit its multilingual generalization ability. Our work follows SSL-based speech anonymization framework, aiming at further improving multilingual performance by integrating and fine-tuning multilingual HuBERT models \cite{zanonboitoMHuBERT147CompactMultilingual2024} as soft content encoders.

\begin{figure}[t]
  \centering
  \includegraphics[width=\linewidth]{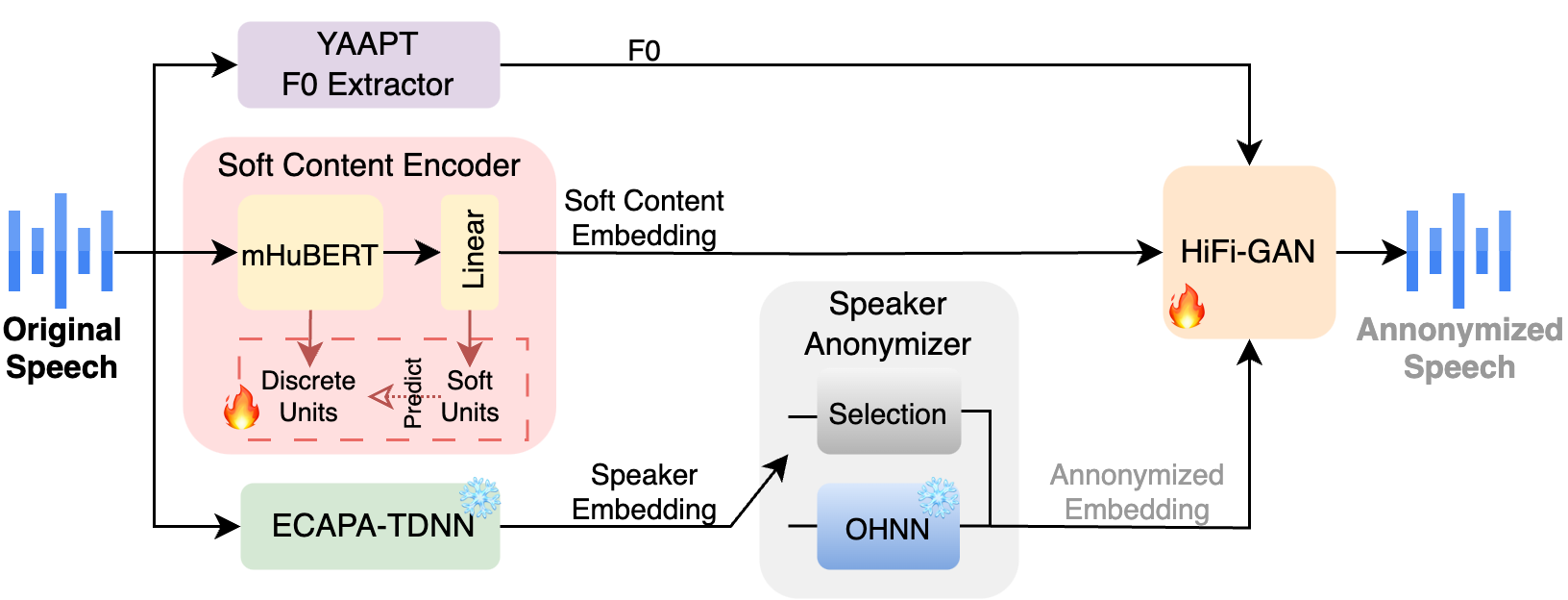}
  \caption{Framework of the SSL-based multilingual SAS.}
  \label{fig:sas}
\end{figure}

\section{Methods}
\label{sec:methods}
In this section, we describe framework of the proposed SSL-based multilingual SAS, which is illustrated in \autoref{fig:sas}. Then, we introduce the details of the multilingual soft content encoder, including the foundation models and fine-tuning procedures.

\subsection{SSL-based Speaker Anonymization System}

Our SAS framework consists of a HuBERT-based soft content encoder, an ECAPA-TDNN \cite{desplanquesECAPATDNNEmphasizedChannel2020} speaker encoder, an F0 extractor, and a HiFi-GAN \cite{kongHiFiGANGenerativeAdversarial2020} vocoder. The soft content encoder is obtained by fine-tuning a pre-trained HuBERT \cite{hsuHuBERTSelfSupervisedSpeech2021} foundation model with an extra layer. The choice of the foundation model and the fine-tuning data is the key part of our research question, which will be discussed in detail. The fine-tuned soft content encoder outputs continuous representations of speech content. The ECAPA-TDNN speaker encoder provides an utterance-level speaker representation. The YAAPT algorithm \cite{kasiAnotherAlgorithmPitch2002} is used to extract F0 from the input utterance. Then, the content features, F0, and anonymized speaker vector are fed into the HiFi-GAN vocoder to synthesize the anonymized speech.

Regarding the choice of anonymizer, we consider two representative anonymization techniques in addition to resynthesis to investigate the effects of SSLs in different settings:
\begin{itemize}
    \item \textbf{Resynthesis with original speaker embedding}: Reconstruct the audio input with original features. The aim is to assess the quality of the vocoder with extracted features.
    \item \textbf{Selection-based anonymizer}: Select an averaged speaker embedding from an external speaker pool, combined with other features to generate anonymized speech.
    \item \textbf{OHNN-based anonymizer}: The OHNN acts like a rotation to transform the original speaker vectors into anonymized speaker vectors, which are constrained to follow the distribution over the original speaker vector space \cite{miaoSpeakerAnonymizationUsing2023}.
\end{itemize}

\subsection{Multilingual Soft Content Encoder}

The HuBERT model is trained by predicting the pseudo labels obtained from an iterative refined clustering \cite{hsuHuBERTSelfSupervisedSpeech2021}, exclusively with English speech data. Aiming at verifying the effects of multilingual pre-training, we investigate a recent multilingual HuBERT model, mHuBERT \cite{zanonboitoMHuBERT147CompactMultilingual2024}, as the basis of our content encoder. We follow \cite{vanniekerkComparisonDiscreteSoft2022} to fine-tune the HuBERT and mHuBERT foundation models with an additional linear layer to produce continuous content representations, which is reported to effectively capture more accurate content information while suppressing speaker information. We first fine-tune a soft HuBERT model with English speech data as a baseline language-independent SAS \cite{miaoLanguageIndependentSpeakerAnonymization2022}. Then, we utilize a Japanese speech dataset to fine-tune both HuBERT and mHuBERT with the same settings. We use the official provided $K$-means models \footnote{\url{https://github.com/facebookresearch/fairseq/tree/main/examples/hubert}} \footnote{\url{https://huggingface.co/utter-project/mHuBERT-147/tree/main}} to generate pseudo labels to train the soft content encoders. After the soft content encoders are trained to convergence, we train the corresponding HiFi-GAN \cite{kongHiFiGANGenerativeAdversarial2020} vocoders for each content encoder to synthesize speech waveforms from the extracted features. Other modules, including the F0 extractor and speaker encoder model, remain the same for fair comparisons. The HiFi-GAN models are trained with the same dataset as that used in fine-tuning the soft content encoders.

To test for language mismatch effects, we compare the above English-only SSL encoder against multilingual SSL encoders in downstream tasks with different scenarios. Specifically, we denote the comparing models as follows:
\begin{itemize}
    \item \texttt{HU-EN}: HuBERT model fine-tuned by English data.
    \item \texttt{HU-JA}: HuBERT model fine-tuned by Japanese data.
    \item \texttt{mHU-JA}: mHuBERT model fine-tuned by Japanese data.
\end{itemize}

\section{Dataset}
\label{sec:dataset}
In the training stage, \textit{LibriSpeech-train-clean-100} dataset is used \cite{panayotovLibrispeechASRCorpus2015} to fine-tune the HuBERT foundation model in \texttt{HU-EN}. For a fair comparison, we randomly sampled 100 hours of Japanese speech from \textit{Corpus of Spontaneous Japanese} dataset \cite{maekawaCorpusSpontaneousJapanese2003} and segmented the speech audio to utterance-level with random durations between 2 and 20 seconds to fine-tune the soft content encoders in \texttt{HU-JA} and \texttt{mHU-JA}. The corresponding HiFi-GAN vocoder models are trained from scratch with the same data. 

In the evaluation of the comparing methods, we conducted two types of downstream tasks, i.e., ASR for utility and ASV for privacy. In addition, we considered two kinds of scenarios: language-adapted and language-expanded. In the language-adapted scenario, we utilized \textit{JVS corpus} \cite{takamichiJVSCorpusFree2019} dataset for ASR experiments, transcribing 9,976 utterances anonymized by each approach. \textit{JTubeSpeech} \cite{takamichiJTubeSpeechCorpusJapanese2021} was adopted for ASV experiments, with 76 enrolled utterances and 276 test utterances creating a total of 20,976 trial pairs for evaluation. In the language-expanded scenario, we followed the setting in \cite{miaoSpeakerAnonymizationUsing2023} to conduct experiments with Mandarin speech. The \textit{AISHELL-3} dataset \cite{shiAISHELL3MultispeakerMandarin2021} was used for both the ASR experiments and ASV experiments. We split the utterances in the test set into test trial (88 utterances) and enrollment (4,179 utterances) subsets, which produced 10,120 enrollment-test pairs for ASV evaluation.

\begin{table*}[th]
  \caption{Evaluation of privacy and utility of SASs in Japanese and Mandarin.}
  \label{tab:eval}
  \centering
\begin{tabular}{lcccccccccc}
\toprule
\multirow{2}{*}{Metrics (\%)} & \multirow{2}{*}{Original} & \multicolumn{3}{c}{Resynthesis} & \multicolumn{3}{c}{Selection-based Anonymizer} & \multicolumn{3}{c}{OHNN-based Anonymizer} \\
\cmidrule(lr){3-5} \cmidrule(lr){6-8} \cmidrule(lr){9-11}
& & \texttt{HU-EN} & \texttt{HU-JA} & \texttt{mHU-JA} & \texttt{HU-EN} & \texttt{HU-JA} & \texttt{mHU-JA} & \texttt{HU-EN} & \texttt{HU-JA} & \texttt{mHU-JA} \\
\midrule
Japanese $\text{EER}_{\text{ja}} \uparrow$ & 14.91 & 29.41 & 27.19 & 24.90 & 47.87 & 48.15 & 39.91 & 43.33 & 44.70 & 39.44 \\
Japanese $\text{CER}_{\text{ja}}^{\text{kata}} \downarrow$ & \phantom{0}3.03 & \phantom{0}5.27 & \phantom{0}4.35 & \phantom{0}\textbf{4.04} & \phantom{0}5.57 & \phantom{0}4.88 & \phantom{0}\textbf{4.18} & \phantom{0}6.18 & \phantom{0}4.78 & \phantom{0}\textbf{4.68} \\
Japanese $\text{CER}_{\text{ja}}^{\text{kanji}} \downarrow$ & \phantom{0}6.94 & \phantom{0}9.58 & \phantom{0}8.37 & \phantom{0}\textbf{8.06} & \phantom{0}9.97 & \phantom{0}9.03 & \phantom{0}\textbf{8.18} & 10.93 & \phantom{0}8.95 & \phantom{0}\textbf{8.90} \\
\midrule
Mandarin $\text{EER}_{\text{cn}} \uparrow$ & \phantom{0}5.56 & 18.76 & 16.51 & 14.09 & 44.33 & 41.62 & 35.21 & 42.55 & 33.28 & 31.20 \\
Mandarin $\text{CER}_{\text{cn}} \downarrow$ & \phantom{0}7.50 & 23.10 & 15.95 & \textbf{10.39} & 25.97 & 21.24 & \textbf{12.67} & 25.74 & 22.76 & \textbf{14.13}\\
\bottomrule
\end{tabular}
\end{table*}

\section{Evaluation}

\subsection{Experiment Setup}

To assess the anonymizer's ability to protect speaker identity, we adopted the privacy metric and evaluation methods from VPC. Specifically, we tested the ASV performance in terms of EER by using an ECAPA-TDNN \cite{desplanquesECAPATDNNEmphasizedChannel2020} model trained on multilingual speech data. ASV experiments are conducted under the \textit{ignorant} \cite{miaoSpeakerAnonymizationUsing2023} setting, where attackers are unaware of the anonymization strategy used for the test trial utterances and use the original enrollment data to infer a speaker’s identity. To assess speech content and intelligibility of anonymized speech, the ASR performance in terms of the CER was computed as a utility metric. We used a state-of-the-art multilingual ASR model, \texttt{whisper-large-v3} \footnote{\url{https://huggingface.co/openai/whisper-large-v3}}, to transcribe anonymized speech and compute CERs with original texts. All the texts were normalized before computing CERs to reduce side effects such as punctuations and number characters.

\begin{figure}[t]
  \centering
  \includegraphics[width=\linewidth]{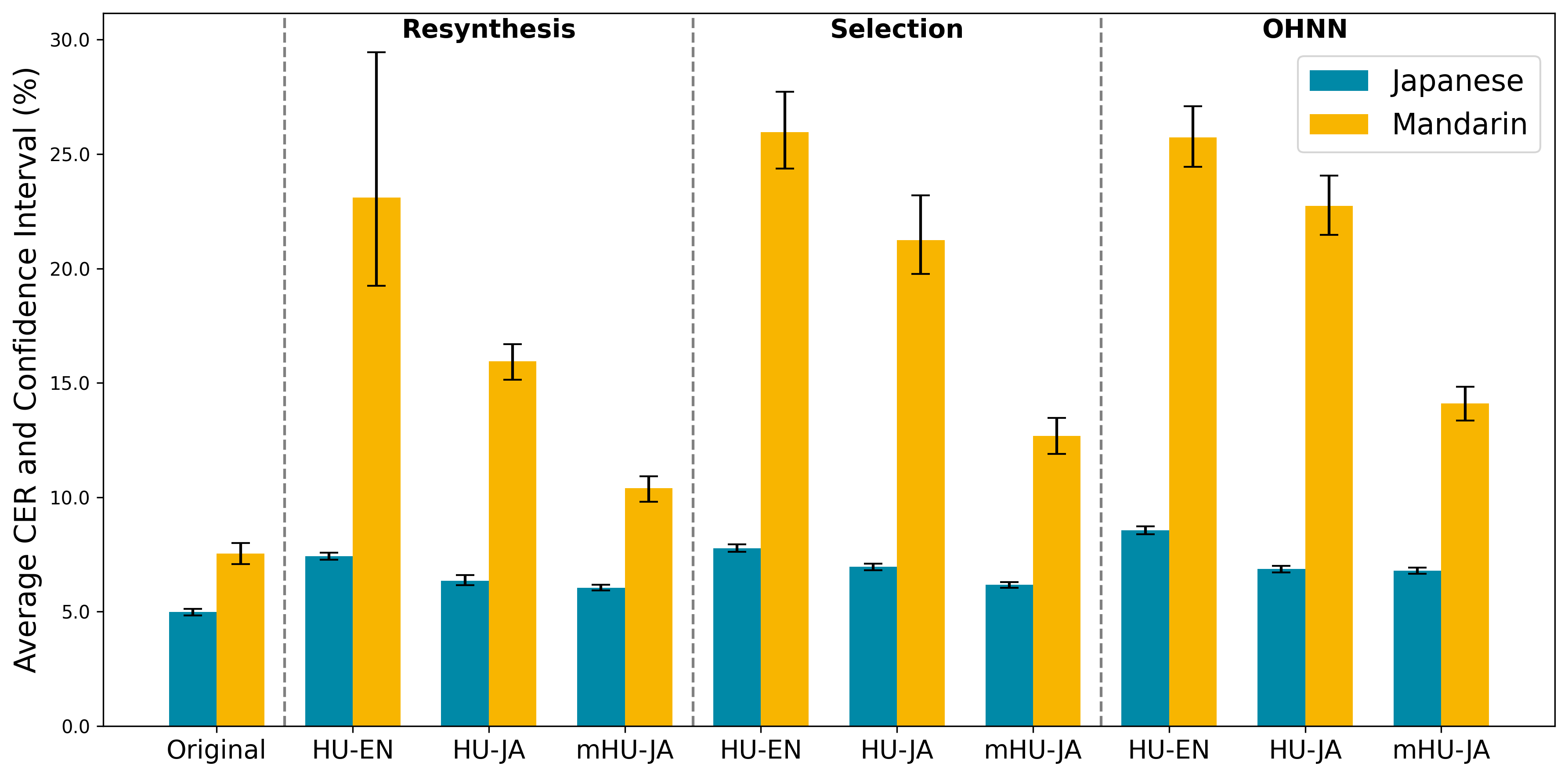}
  \caption{CER scores and confidence intervals.}
  \label{fig:confidence_intervals}
\end{figure}

We evaluate the SASs in two kinds of scenarios: language-adapted and language-expanded. Since our proposed models are fine-tuned on Japanese speech, we first evaluate the SASs with a Japanese test set to confirm the effectiveness of fine-tuning. Moreover, we evaluate the SASs with another test set in Mandarin, which is an unseen language in SSL fine-tuning and vocoder training. This targets assessing whether the multilingual SSL still maintain the linguistic scope after fine-tuning with a specific language. We anonymized the utterances in the test sets using different SASs described in \autoref{sec:methods}. Then we compared the evaluation metrics from different models. We also include the ground-truth utterances to confirm the performance of the evaluation pipeline and the ideal upper bound of the metrics, as shown in the \textit{Original} column in the \autoref{tab:eval}. \textit{Resynthesis} with extracted features unchanged is also tested as a basic scenario, which can produce higher EERs in ASV than \textit{Original} because of vocoder drifts \cite{panarielloVocoderDriftXvectorbased2023}. In addition, we conducted statistical significance test with bootstrapping approach \footnote{\url{https://github.com/luferrer/ConfidenceIntervals}} on ASR results and present the confidence intervals in \autoref{fig:confidence_intervals}.

\subsection{Language-Adapted Evaluation: Japanese}
In the language-adapted case, the ASV results of Japanese speech are denoted as $\text{EER}_{\text{ja}}$ in \autoref{tab:eval}. Japanese texts have two different representations: \textit{Kanji} and \textit{Katakana}. We present both levels as $\text{CER}_{\text{ja}}^{\text{kata}}$ and $\text{CER}_{\text{ja}}^{\text{kanji}}$. Compared to Katakana-level texts representing the pronunciation units (syllables), Kanji-level transcription can be more difficult because one Katakana character combination may correspond to different Kanji characters, hence requiring more subtle voice details from the prosody, accent, and contexts. The ASV results confirmed the effective privacy preservation of SASs. While defending against the attackers with high EERs, the CERs from ASR tasks consistently decreased when using the soft content encoders fine-tuned with Japanese speech (\texttt{HU-JA} and \texttt{mHU-JA}) compared to the original English-based SSL model (\texttt{HU-EN}). This verified our hypothesis that integrating a language-adapted SSL model can improve the utility of SASs while preserving privacy. Moreover, the multilingual models (\texttt{mHU-JA}) can further improve the intelligibility. We argue that the English-only HuBERT model can have difficulty retrieving the exact phoneme representations in other languages, resulting in accent and pronunciation mismatches in the anonymized speech. 

\subsection{Language-Expanded Evaluation: Mandarin}
As an out-of-domain language in the fine-tuning phase, the Mandarin test dataset was also utilized as a language-expanded scenario for evaluating the fine-tuned multilingual SSL. We first confirmed reasonable EER scores from ASV experiments, verifying that the anonymizers still protect the speaker identity successfully in Mandarin case. While the ASR results for \texttt{HU-EN} show a significant deterioration in intelligibility for Mandarin, the \texttt{HU-JA} models already demonstrate an improvement, even though Mandarin remains a zero-shot language for both. Fine-tuning on Japanese equips \texttt{HU-JA} with a representation space that captures phonetic features, such as overlapping phonemes and similar syllable structures. This acoustic foundation allows the model to retrieve more accurate content representations, resulting in anonymized speech that better preserves natural pronunciation and accent nuances. Most importantly, \texttt{mHU-JA} significantly decreases the CERs of Mandarin, verifying our hypothesis that multilingual SSL models can improve the utility of anonymized speech data while maintaining the privacy preservation of the anonymizers in an expanded linguistic scope.

\begin{figure}[t]
  \centering
  \includegraphics[width=\linewidth]{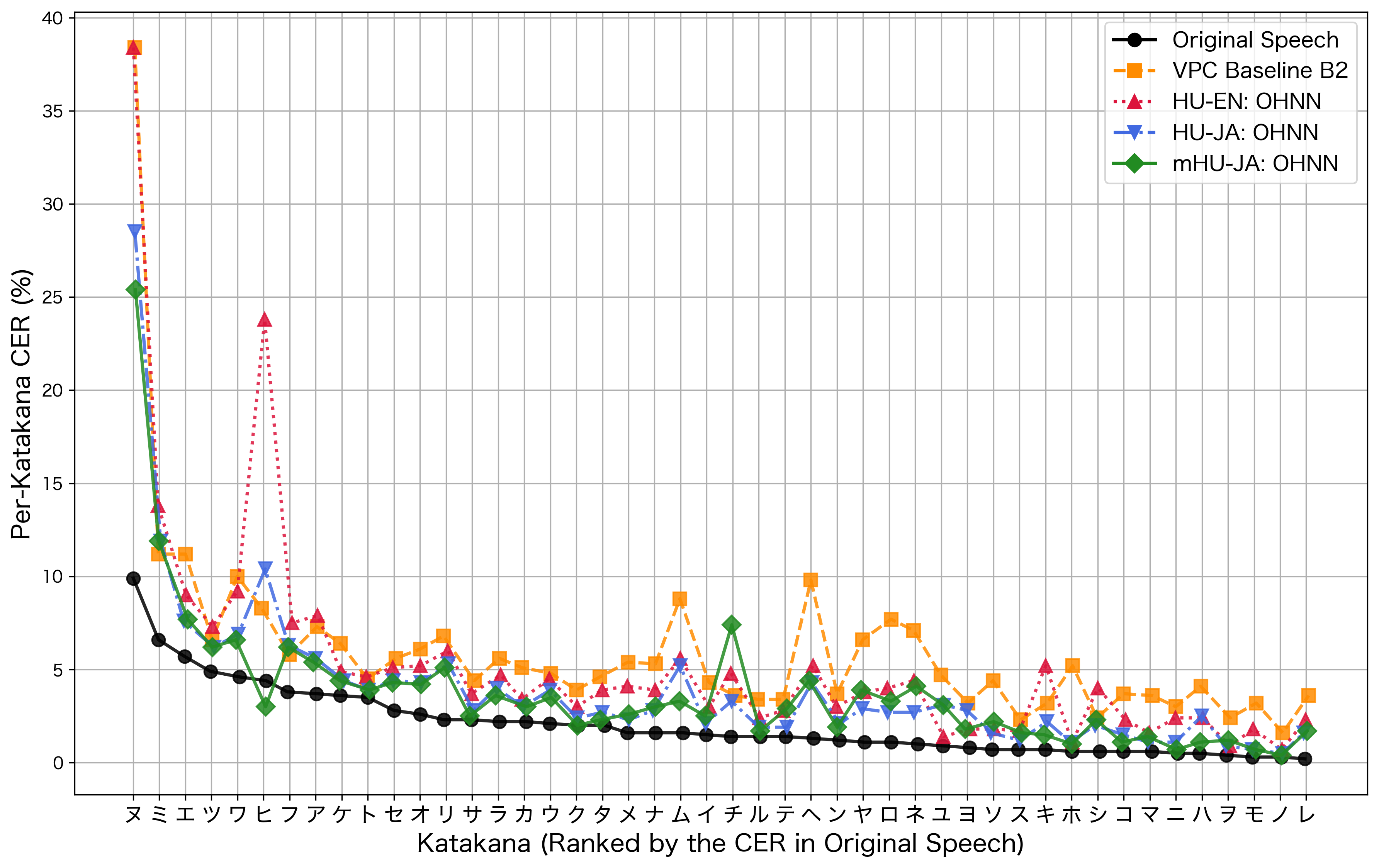}
  \caption{Per-Katakana CER of synthesized Japanese Speech.}
  \label{fig:per_katakana_cer}
\end{figure}

\subsection{Phonetic Analysis}
To further analyze the effects of SSL-based encoders, we computed per-Katakana CER of Japanese speech. Katakana is a syllabary representing phonetic blocks in Japanese speech. We take \textit{OHNN} anonymizer as a representative for our proposed methods, as shown in \autoref{fig:per_katakana_cer}. The fine-tuned multilingual SSL (\texttt{mHU-JA}) consistently demonstrates better intelligibility. For some specific Katakana, such as \begin{CJK*}{UTF8}{min}ヌ\end{CJK*} (NU) and \begin{CJK*}{UTF8}{min}ヒ\end{CJK*} (HI), the English-based model (\texttt{HU-EN}) presents a poor performance, while our proposed SASs can significantly improve the recognition accuracy, thus improving the utility of anonymized speech.

\subsection{The Privacy-Utility Tradeoff}
Despite successfully confirming the effectiveness of multilingual SSL models, we notice that the EER of ASV results slightly decrease when using the \texttt{HU-JA} and \texttt{mHU-JA} models. This is because the better intelligibility of anonymized data also makes it easier for the ASV system to identify the speaker. This issue is interpreted as a \textit{privacy-utility tradeoff} in the VPC report \cite{tomashenkoVoicePrivacy2024Challenge2024} to address the situation that some SASs are configured to offer better privacy at the cost of utility and vice versa. Overall, our proposed methods achieve a good utility-privacy tradeoff by exploiting multilingual SSL models.

\section{Conclusion}

In this paper, we presented an extended investigation of language mismatch in SSL-based speaker anonymization. Our study revealed that English-only SASs lead to significant utility degradation when processing Japanese and Mandarin speech. By fine-tuning content encoders and incorporating multilingual SSL models, we substantially improved speech intelligibility in both language-adapted and language-expanded scenarios while maintaining robust speaker privacy. Our experiments in ASR and ASV tasks demonstrated that language adaptation and multilingual pre-training are keys to mitigating the language mismatch in SASs. The improvements observed in CER and EER metrics highlighted the potential of these approaches for real-world applications in a wider linguistic scope.

Future work will extend this investigation to additional languages and dialects and examine adaptation strategies. In addition, in-depth analyses through phonetic and linguistic studies promise to reveal more meaningful insights from the experiments. Moreover, we plan to conduct listening tests to further assess the SASs with naturalness and perceived anonymity.

\section{Acknowledgments}
This study is partially supported by JST AIP Acceleration Research (JPMJCR24U3) and by MEXT KAKENHI Grants (24K21324). This study was carried out using the TSUBAME4.0 supercomputer at Institute of Science Tokyo.

\bibliographystyle{IEEEtran}
\bibliography{mybib}

\end{document}